Data Integration and spatio-temporal statistics can quantify relative risk of medico-legal reforms: the example of police emergency mental health responses in Queensland (Australia)

Nidup Dorji[a,*,1], Sourav Das[a,*,2], Richard Stone[c] and Alan R Clough[b,3]

[a]School of Electrical Engineering, Computing, and Mathematical Sciences, Curtin University, Perth, WA, Australia
[b]Professorial Research Fellow, College of Medicine and Dentistry, and Australian Institute of Tropical Health and Medicine, James Cook University, PO Box 6811, Cairns, QLD, Australia
[c] Director Emergency Medicine, Cairns Hospital, Cairns, QLD, Australia

ABSTRACT

This study examined the spatial-temporal dynamics of Emergency Examination Order/Authority (EE-O/A) admissions in Far Northern Queensland (FNQ) from 2009 to 2020, using 13,035 unique police records aggregated across 83 postcodes. A two-stage modelling framework was applied: Lasso regression identified a parsimonious set of socio-economic and health-service covariates, and a Bayesian Conditional Autoregressive (CAR) model incorporated these predictors with structured spatial and temporal random effects. Exploratory analysis revealed a general upward trend in annual admissions, with a pronounced peak in 2017 (1,305 cases) and Moran's I statistics indicated significant positive spatial autocorrelation after a 2017 policy change (Global Moran's I = 0.138–0.151, $p < 0.05$). The CAR model indicated that higher proportions of Aboriginal (Indigenous) residents (posterior mean=0.351, 95% CI 0.235–0.463) and greater pharmacy density (posterior mean=0.422, 95%CI 0.174–0.672) were positively associated with EE-O/A counts, while average income (posterior mean=-0.697, 95%CI-1.196 to-0.205) and white-collar occupation (posterior mean=-0.537, 95%CI-0.695 to-0.378) showed negative associations. Spatial risk maps highlighted clusters of elevated relative risk around postcodes 4870 (Cairns), 4740 (Mackay) and 4810 (Townsville), i.e. FNQ's major cities. The findings suggest that socio-economic disadvantage and service accessibility drive EE-O/A incidence, underscoring the need for targeted mental-health interventions and resource allocation in impoverished FNQ communities. Limitations include reliance on cross-sectional census data for covariates and potential ecological bias from data fusion.

## 1. Introduction

Emergency departments (EDs) are critical points of care for people experiencing acute mental health crises, including those subject to state-authorised, involuntarily transport from the community by police or ambulance services [1, 2]. In Queensland, police and ambulance officers may detain and transport a person under the Emergency Examination Authority (EEA) provisions of the *Public Health Act 2005*(Qld) (PHA) to a public sector health service facility in emergency circumstances [1, 3]. The provisions are triggered if the officer has reason to believe: a) a person's behaviour indicates that the person is at immediate risk of serious harm; (b) the risk appears to be the result of a major disturbance in the person's mental capacity, whether caused by illness, disability, injury, intoxication or another reason and (c) the person appears to require urgent examination, or treatment and care, for the disturbance. The PHA specifies the example of threatening to commit suicide as indicative of immediate risk of serious harm. These legislative provisions entered into force on 5 March 2017.

Before this, police and ambulance services were authorised to detain and transport a person under the Emergency Examination Order (EEO) provisions of the *Mental Health Act 2000* (Qld) (MHA) – now repealed – if the officer had a reason to believe that the person was at risk because they may have had a mental illness. Our study enumerates episodes using Queensland Police Services (QPS) records and does not explicitly differentiate between an EEO or EEA. Hence, the term 'EE-O/A' is used throughout to describe the records of these episodes.

*Corresponding author
nidup1998@gmail.com (N. Dorji);
sourav2.das@googlemail.com
(S.Das)

Whether under an MHA EEO prior to 5 March 2017, or a PHA EEA since, in the vast majority of cases, persons transported by police or ambulance are taken to a hospital ED for examination, treatment or care. Unlike for all EEOs prior to 5 March 2017, data facilitating public scrutiny of all EEAs, has not been publicly available since 5 March 2017 and, accordingly, there are few systematic studies of trends and patterns in their occurrence. Mental health crises have led to catastrophic outcomes in Queensland, where crisis situations have escalated to require police intervention with deadly force, as happened in a cluster of shootings by police in south-east Queensland in 2013-14[4] or a detained person dies in custody, as in the tragic circumstances where an Indigenous man died in an ambulance before reaching the ED of a far north Queensland hospital in 2018 following a physical struggle with police[5].

Given the identified evidence gap, and recognising the very high stakes nature of EE-O/A episodes, this research has the following objectives:

(a) investigate systematic public health factors that have driven police-registered EE-O/A events in FNQ based on data-integration of requisite but disparate socio- economic, clinical, and economic information;

(b) explore trends in spatial clustering of the number of episodes at postcode level employing state-of-

the- art, probabilistic spatial modelling to obtain spatial relative risk; and

(c) use the model to detect potential spatial and temporal heterogeneities in public health responses.

The article is organised as follows. In Section 2, we introduce the data and associated data fusion pipeline that was used to assimilate various potential exposure to and EE-O/A, Section 3 defines all related statistical methods including unsupervised and supervised algorithms, Section 4 describes observations and inferences. We conclude with discussion and limitations in Section 5.

## 2. Data

### 2.1. Setting

The study was conducted in the Far Northern region of Queensland (FNQ), encompassing both remote and non-metropolitan areas of the state together with the major urban centres in FNQ. The region represents approximately 15% of Queensland's total population of 5.62 million people [6].

### 2.2. Sampling

A total of 13,035 unique EE-O/A records were registered by QPS in the region's four Hospital and Health Services (HHS) districts, from 2009 to 2020. Geographically, the spatial occurrence of these records spans 83 of the 447 postcodes of Queensland, reflecting a substantial geographic and demographic spread of EE-O/A admissions across the region.

### 2.3. Data Sources

Additionally, for developing a spatio-temporal supervised model to be able to explain the epidemiology of EE-O/As, we also sampled and integrated the following likely concomitant-confounding explanatory features that are reported in disparate public or web-based repositories. All information was subsequently aggregated at Queensland's postcode level. For any missing values, we used univariate probabilistic imputation where applicable [7].

  i. **EEA-O/A data**: Patient-level data (demographic information related to patients transported to an ED) from 13,035 EE-O/A records generated by QPS. Since 2009, the "Queensland Police Records and Information Management Exchange" (QPRIME) has finalised all EE-O/A incidents according to consistently maintained statewide QPS protocols to ensure quality of records for enforcement and investigation using categories derived from legislative provisions.
  ii. **Socioeconomic Indicators**: Data on demographic, income, and employment status were sourced from 2021 census data, Australian Bureau of Statistics [6].
  iii. **Education and Public Services**: Data on schools and libraries were retrieved from the Australian Government Open Data Portal [8].
  iv. **Pharmacy**: Data on pharmacies, including locations, were sourced from the Department of Pharmacy Business Ownership Administrative System (PBOAS), Queensland Health [9].
  v. **General Practitioner**: Data on general practitioners (GP)s sourced from the Australian Commission on Safety and Quality in Health Care [10] using Python Script.
  vi. **Geospatial Boundaries**: Shapefiles delineating 447 postal regions in Queensland were also obtained from the Australian bureau of statistics [6].

Appendix A shows detailed data collection, data processing and cleaning, and data assimilation pipeline at a postcode level.

## 3. Methodology

The key probability models of supervised learning used to model spatially varying relative risk of annual EE-O/A incidents observed in this data and its uncertainty are described as follows.

### 3.1. Preliminaries

Define, $Z(s_i, t_j)$, to be the number of EE-O/A cases reported in postcode $s_i$ in year $t_j$ resulting in the observation domain $A = \{(s_i, t_j), i = 1,2, \ldots, N; j = 1,2, \ldots, n\}$ with mean $\mu(s_i, t_j) = \mathbb{E}Z(s_i, t_j)$, and spatial variance $\Sigma_s = \mathbb{C}\text{ov}(Z(s_i, t_j), Z(s_k, t_l))$. A key objective of statistical modelling would be parametric estimation of $\mu(s_i, t_j)$ and $\Sigma_s$, explaining spatio-temporal dynamics of EE-O/As.

### 3.2. Spatial clustering

Just as in case of a linear modelling, we begin by considering exploratory evidence for spatial variation. A statistical measure of clustering or spatial heterogeneity over lattice sampling regions (such as postcodes) was defined by Moran [11]. Moran's measure is a spatially adaptive sample correlation function defined as follows.

$$I = \frac{n}{(n-1)S^2 w..} \sum_{i=1}^{N} \sum_{j=1}^{N} w_{ij}(Z(s_i) - \bar{Z}_1)(Z_1(s_j) - \bar{Z}_1)$$

$$S^2 = \frac{1}{n-1} \sum_{i=1}^{N} (Z_1(s_i) - \bar{Z}_1)^2$$

$$w\_\{ij\} = \begin{cases} 1, \text{if postcodes } i \text{ and } j \text{ are connected} \\ 0, \text{if postcodes } i \text{ and } j \text{ are connected} \end{cases}$$

$$w.. = \sum_{i=1}^{N} \sum_{j=1}^{N} w_{ij}$$

Moran's cluster statistic $I1$ can have positive (clustered), negative (dispersed, repulsive) or values close to zero (complete random patterns). Note that among its limitations is the fact that it is unable to distinguish between clustering due to systematic or stochastic factors. Various modifications to the basic Moran's test have been proposed in the literature, including the local-Moran's test that we implement in R [12]. Interested readers are referred to [13].

In this article, however, our use of Moran's test is exploratory and indication of spatial autocorrelation of EE-A/O, thus we treat the problem of explanation of any potential clustering to a spatio-temporal statistical model.

### 3.3. Regularised regression modelling

LASSO [14] is a regression-with-regularisation methodology that improves regression methods for a response $(Z(s_i, t_j)$, say) on high dimensional predictors, say with $p(> Nn)$, predictors, $X(s_i, t_j) = \{X_1(si, tj), X_2(Si, tj), ... X_p(s_i, t_j); i = 1, ..., N, j = 1, ..., n\}$ where the feature set often has potential multi-collinearity. This can impact a supervised learner by making it susceptible to false positives but also over-fitting leading to higher mean squared errors (MSE). LASSO penalised the magnitude of coefficients incurring higher bias but lower MSE, that is the key accuracy parameter in out-of-sample validation. If

$$\{Z(s_i, t_j) \mid X(s_i \sim f_Z(z(s_i, t_j); \boldsymbol{\beta}),$$

an exponential family distribution, independently,

$$\mathbb{E}[g(Z(s_i, t_j) \mid X((s_i, t_j)\}] = X((s_i, t_j)\boldsymbol{\beta},$$

the link function, LASSO estimates the coefficients of the linear model by maximising

$$\underset{\beta}{argmax}\, Q_p = \sum_{i=1}^{N}\sum_{j=1}^{n} log f_Z(z(s_i, t_j); \boldsymbol{\beta}) - \lambda \sum_{q=1}^{p} |\beta_q|,$$

$\lambda$, the penalty is estimated using cross-validation. We implement LASSO in R using the package `glmnet` ([15]). note that model 6 only accounts for systematic variation assuming spatial independence. In this paper we use LASSO for selecting a subset of systematic explanatory factors that associate with EE-O/A incidence before proceeding with a complete spatio-temporal model.

### 3.4. Model for spatial variation

We assume that the spatio-temporal variation in $Z(s_i, t_j)$ can be decomposed into large scale or environmental factors $X(s_i, t_j)$ and stochastic variation. $X(s_i, t_j)$ may include factors such as income, literacy, age, pharmaceutical availability, leisure, and other welfare support infrastructure. Stochastic variation accounts for probabilistic spatio-temporal dependency of $Z(s_i, t_j)$ on its neighbours. For more details on the spatio-temporal framework of spatio-temporal modelling, see [16, 17].

#### 3.4.1 Neighbourhood

Let $N_{ij}$ denote the spatio-temporal neighbourhood set of the observation tuple $(s_i, t_j)$ with $N_{ij}$ the neighbour-hood size, demonstrated in Figure 1. $N_{ij}$ accounts for spatio-temporal dependency of a given location-year sample $Z(s_i, t_j)$ on its neighbours. In other words, non-neighbour location-year tuples do not influence $Z(s_i, t_j)$. For example, if the spatio-temporal neighbour set for postcode 1 in year 5, $Z(s_1, t_5)$ is $N_{15} = \{(s_3, t_4), (s_7, t_4), (s_8, t_4)\}$ only year $t_4$ of postcodes $(s_3, s_7, s_8)$ affect $Z(s_1, t_5)$. Other variation in $Z(s_1, t_5)$. would be assumed to be environmental that include demographic and other socio-economic factors. [18] following on from[19] and statistical modelling of time series defined a spatio-temporal probability *Markov* process for lattice data $Z(s_i, t_j)$. Akin to time-series $Z(s_i, t_j)$ has a probability distribution that is only dependent on immediate neighbourhood. Besag's definition of spatial neighbours in $N_{ij}$ would imply that postcodes are neighbours only if they are adjacent (sharing spatial boundaries). This is directly related to the first principle of geography ([20]) proximate locations lead to similar event records, other similarities are accounted for by systematic variation $(X(s_i, t_j)$

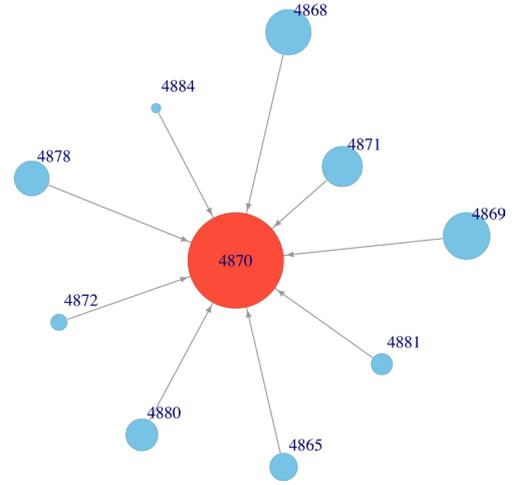

**Figure 1:** Neighbourhood structure: The plot uses graphical network architecture to depict spatial dependency of a major EE-O/A postcode (4870) on its neighbours. The radii of the nodes are proportional to the relative risk (RR) 7.

#### 3.4.1. Conditional autoregression

The statistical purpose of a spatio-temporal model is to be able to construct the multivariate joint distribution $f\{z(s_1, t_1), (s_1, t_2), ..., z(s_n, t_N)\}$, the complete description of spatio-temporal dynamics [17], comprising systematic is of comprising systematic and stochastic effects. But the full joint distribution is often complex and intractable. However, under the above framework, and conditions of Hammersley-Clifford theorem Besag showed that [18, 21], the conditional spatial-temporal autoregressive (CAR) model for any multivariate distribution $f(.)$ describes the full joint relationship between $Z(s_i, t_j)$, systematic variables $(X(s_i, t_j)$, and its neighbourhood $N_{ij}$.

$$\{Z(s_i, t_j) \mid S(s_i, t_j)\} \sim f_c\{n(s_i, t_j)\lambda(s_i, t_j), \Sigma_{CAR}\};$$

$$\lambda(s_i, t_j) = E\{Z(s_i, t_j) \mid S(s_i, t_j)\} \text{ and}$$

$$Var\{Z(s_i, t_j) \mid S(s_i, t_j)\} = \Sigma_{CAR} \text{ (say)}$$

$n(s_i, t_j)$: an appropriate offset usually susceptible

$$\log\{\lambda(s_i, t_j)\} = \boldsymbol{X}(s_i, t_j)^t \beta + S(s_i, t_j) \text{population}$$

$$\boldsymbol{S}(N) \sim G(\boldsymbol{0}, \Sigma_C(\rho))$$

$g(.)$ is link function. $c_{ij}$ defines the stochastic spatial dependence and is non-zero only if an observation is part of the defined neighbourhood $(s_l; t_k) \in N_{ij}$. A key advantage of the conditional modelling approach described in equations 7 is that under the celebrated Hammersley-Clifford theorem it defines a valid multivariate joint distribution with the same mean $X\beta$ and joint covariance matrix

$$\Sigma_{CAR} = (I - C)^{-1}\Sigma_C; \Sigma_C = \text{diag}(\sigma_1^2, \sigma_2^2, \ldots, \sigma_n^2 N)$$

$\boldsymbol{C}$ is spatial neighbourhood (or adjacency) matrix whose entries are 1 (neighbours) or 0 (non-neighbours), corresponding to a pair of postcodes. In case of annual EE-O/A incidence $Z(s_l; t_j)$ observations we considered the Poisson and Negative Binomial distributions, as models for condition density. $f_c\{n(s_i, t_j)\lambda(s_i, t_j), \Sigma_{CAR}\}$ uniquely describes the complete joint probability distribution. For para-meter estimation and inference of the proposed spatio-temporal CAR as shown in Equation 7, we used the integrated-nested- Laplace transform (INLA) [22] that is more efficient than conventional Markov Chain Monte-Carlo (MCMC) based approaches. INLA was implemented using the package INLA [23] in R. INLA also implements the neighbourhood architecture described and offers alternatives. Further details on INLA as a procedure for statistical estimation in different applications can be found in [24, 25]. We use model Equation 7 to estimate a). the spatio-temporal variation of the relative risk of the incidence of EE-O/A [26], relative to completely random spatial occurrence b). but we also estimate the relative risk compared to the administrative region with the lowest risk (postcode).

## 4. Results

### 4.1. Exploratory Analysis and Spatial Pattern

Figure 2 shows the annual admissions to EDs under EE-O/As from 2009 to 2020. Despite some troughs, we observe a general increasing temporal trend in reported EE-O/A cases. The highest number of admissions was recorded in 2017 (1305), the year the amended PHA entered into force with the regulatory definitions no longer restricted to mental illness but broadened to encompass the notion of imminent risk of serious harm, such as a threat to commit suicide. The 2017 figure of 1305 represents an approximate 20% increase over 1085 EE-O/As the previous year.

Choropleth maps of average admissions (Figure 3a) and standard deviation (Figure 3b) of EE-O/A admissions from 2009 to 2020 in each postal region reveal a spatial variation in EE-O/A admissions. The spatial variation of the average admissions (Figure 3a) indicates comparable incidence rates among the neighbouring postal regions, suggesting patterns could be associated with, for example, shared environmental

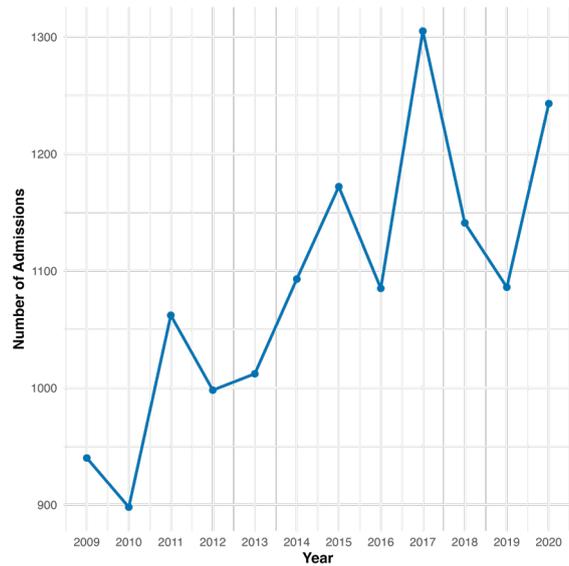

**Figure 2:** The line graph represents the trend of annual EE-O/A admissions in far northern Queensland, covering 83 postcodes, from 2009 to 2020.

, socio-economic, demographic, health, and leisure services factors.

The distribution of covariates including demographic (Aboriginal population; Figure 4a), socioeconomic (average income; Figure 4b), retail (liquor outlets; Figure 4c), and health services (general practice clinics; Figure 4d) showed similar spatial variation between postal regions. The observed spatial patterns indicate a potential systematic association between the underlying socioeconomic, environmental, and demographic factors and EE-O/A admissions across the study regions. This is further evidenced by the positive association between EE-O/A admissions and potential confounders observed in the five postal regions with the highest aggregate admissions between 2009 and 2020, shown in Appendix B. Postcode 4870 (Cairns) by far recorded the highest number of EE-O/A cases ($n = 1951$) during the study period. The region also had the highest concentration of liquor stores ($n = 341$), and ranked second in the number of people without income ($n = 3736$), the Aboriginal population ($n = 3,422$), and the individuals employed in blue-collar occupations ($n = 15,247$). Spatial patterns and clustering of the EE-O/A occurrences imply that the EE-O/As at the postcode level are likely influenced not just by the local conditions but also by the characteristics of the neighbouring postal regions, thus requiring further investigation.

### 4.2. Spatial Autocorrelation - Moran's I: Unsupervised learning

Table 1 presents a spatial autocorrelation analysis of admissions under EE-O/A from 2009 to 2020 using Global Moran's I statistics (Equation 1). The Global Moran's I values ranged from 0.023 to 0.164 and the Z-score from 0.485 to 2.370. During the years prior to the policy transition, significant spatial autocorrelation was observed, as indicated by the significance value (p-value) for each year, with the exception of 2009 (Moran I = 0.164, Z score = 2.370, and P-value = 0.009), which showed a significant positive spatial autocorrelation. In contrast, the years following the policy change (2017-2020) consistently demonstrated statistically significant positive spatial autocorrelation. Furthermore, the Global Moran's I statistic calculated for aggregated EE-O/A admissions from 2009 to 2020 across postcodes revealed significant spatial autocorrelation (Moran's I = 0.151, Z-score = 2.412, p = 0.0079), as shown in Table 1. A positive spatial autocorrelation across the 83 postal regions suggests that the clustered patterns of EE-O/A occurrences are unlikely to be due to randomness, indicating a degree of spatial dependency of EE-O/A admissions.

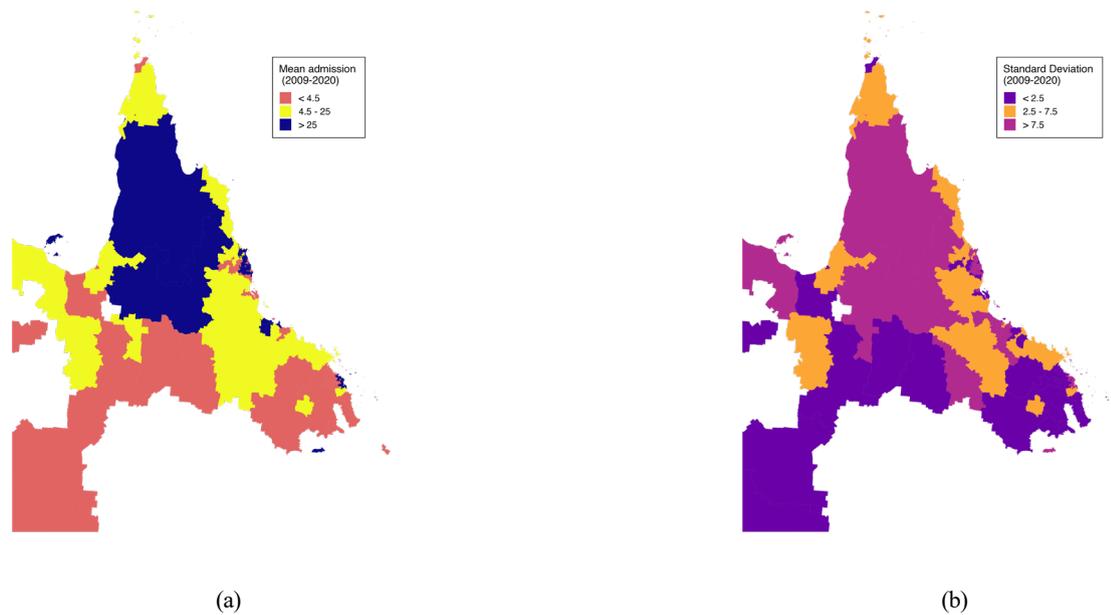

**Figure 3:** a). The choropleth map shows the spatial variation of mean EE-A/O admissions for each postcode from 2009 to 2020, b) and the corresponding standard deviation of EE-A/O admission for each postal region from 2009 to 2020.

### 4.3. Supervised Learning
#### 4.3.1. Feature Selection Using Lasso Regression with Negative Binomial Distribution

Lasso regression, adapted for negative binomial (NB) distribution, was used to select a subset of covariates potentially associated with EE-O/A admissions, while accounting for the overdispersion of the EE-O/A counts (Equation 6).
The variables, (refer to Appendix C for detailed interpretations of variables), were selected using the regularisation parameter *lambda* ($\lambda$), determined based on the one-standard- error rule during cross-validation, resulting in a parsimonious and interpretable model. Table 7, in Appendix D, shows estimated coefficients of the covariates selected by the Lasso-NB model. Notably, demographic variables such as the proportion of Aboriginal, proportion of population with- out formal schooling, individuals engaged in white collar occupations, along with retail indicators such as liquor store count and pharmacy count, were positively associated with EE-O/A counts. In contrast, average income showed negative associations with EE-O/A. Temporal effects (i.e., year-specific variations) were incorporated separately within the CAR modelling framework.

**Table 1**
The table presents the global Moran's I statistic for 13,035 distinct EE-A/O admissions aggregated at the postcode level, along with the yearly Moran's I statistics 1 for EE-A/O admissions from 2009 to 2020.

| Year | P-value | Z-score | Moran's I |
|---|---|---|---|
| aggregated | **0.0079** | 2.412 | 0.151 |
| 2009 | **0.009** | 2.370 | 0.164 |
| 2010 | 0.243 | 0.696 | 0.046 |
| 2011 | 0.314 | 0.485 | 0.023 |
| 2012 | 0.282 | 0.577 | 0.029 |
| 2013 | 0.057 | 1.582 | 0.113 |
| 2014 | 0.171 | 0.950 | 0.061 |
| 2015 | 0.207 | 0.817 | 0.052 |
| 2016 | 0.203 | 0.832 | 0.055 |
| 2017 | **0.025** | 1.959 | 0.138 |
| 2018 | **0.036** | 1.803 | 0.107 |
| 2019 | **0.020** | 2.050 | 0.112 |
| 2020 | **0.040** | 1.751 | 0.105 |

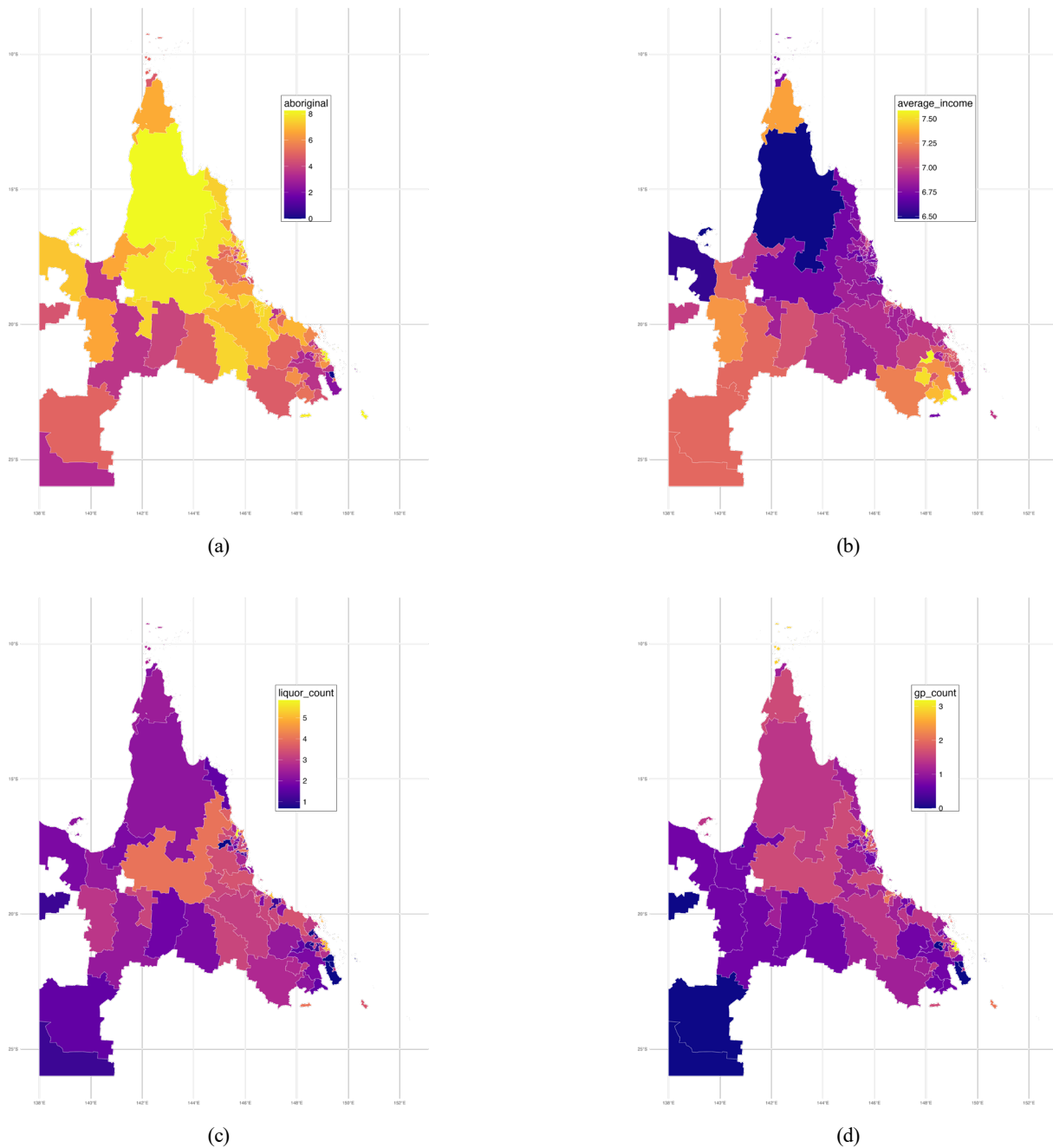

**Figure 4:** a) The choropleth illustrates the logarithmic scaled distribution of the aboriginal population across postcodes based on 2021 ABS census data, b) estimated weekly income distribution9, c) spatial variation of scaled retail liquor outlets, and d) the distribution of the general practitioners.

### 4.3.2. Spatial and temporal Modelling

Hierarchical spatial and temporal models were compared to the Lasso-selected covariates (shown in Table 2), with negative binomial response distribution to account for overdispersion of the EE-O/A counts. **Model 1** considered just fixed effects while **Model 2** added structured spatial random effects for 83 postal regions along with fixed effects. **Model 3** accounted for fixed, spatial, and temporal effects, capturing variation for a period of 13 years. Among the three models evaluated, Model 3, which accounted for fixed effects, spatial and temporal effects, with lowest DIC value of 4038 was the preferred model.

The selected conditional autoregressive spatial and temporal model, **Model 3**, was then fitted to the data, and the co-variates were examined for a potential association with EE- O/A admissions (Equation 7). All covariates retained from the Lasso-NB selection process whose 95% percentile (be- tween 2.5% and 97.5%) did not overlap zero were considered significant. Table 3 illustrates the parameter estimation of significant covariates and effect on EE-O/A counts with 95% credible interval displayed. EE-O/A count was found to be positively associated with the proportion of the Aboriginal population (posterior mean = 0.351, 95% CI: 0.235–0.463) and the number of pharmacies (posterior mean = 0.422, 95% CI: 0.174–0.672). These results indicate that the number of EE-O/As increased with an increase in the number of Aboriginal population and pharmacy in a postcode. The observed association between EE-O/As and pharmacy outlets may be attributed to the concentration of pharmacies in urban areas, where both population density and service accessibility are typically higher compared to rural regions. Other covariates, including the proportion of people without formal education and the number of liquor stores, exhibited positive associations with EE-O/A admissions; however, these associations were not statistically significant, as their 95% credible intervals included zero (see Appendix D for detailed parameter estimation for all NB-selected variables).

**Table 2**
The table shows different model variants fitted to the dataset, along with their corresponding marginal log-likelihood and deviance information criterion (DIC) values for model selection.

| Model | Effects | DIC | Marginal Log-likelihood | Dbar |
|---|---|---|---|---|
| Model 1 | Fixed | 4336 | −2256 | 4210 |
| Model 2 | Spatial | 4329 | −2278 | 4202 |
| Model 3 | Spatial & Temporal | 4083 | −2341 | 3559 |

EE-O/A admissions were found to be significantly and negatively associated with average income, with a mean posterior estimate of −0.697 and a 95% credible interval of (−1.196, −0.205). Although the Lasso-NB feature selection process initially identified white-collar occupation as a potentially positive predictor of EE-O/A admissions, the subsequent conditional regression model confirmed a significant negative association, yielding a mean posterior estimate of −0.537 and a 95% credible interval of (−0.695, −0.378). This indicates that a one-unit increase in average income or white-collar occupation is likely to lead to a decrease in the number of EE-O/A admissions in a postal region by [can you specify a number?]. The reversal in the direction of the association between white-collar occupation and EE-O/A admissions in the CAR model, from positive to negative, may be ascribed to its positive correlation with average income. Given that average income is negatively associated with EE-O/A admissions, it is likely that the effect of white-collar occupation was suppressed when both variables were included in the model.

Figure 5a and Figure 5b present choropleth maps depicting the mean relative risk (RR) and standard deviation of relative risks from 2009 to 2020 in 83 postal regions. Notably, the higher risk areas were clustered around a few postcodes such as 4870 (Cairns), 4740 (Mackay), and 4810 (Townsville), consistent with the spatial and clustering patterns observed in the exploratory analysis. A similar pattern was observed compared to the baseline postcode 4865, which had EE-O/A admissions closest to the average among the 83 postal regions analysed.

## 5. Discussion

This study investigated an enduring medico—legal public health challenge under Queensland's laws authorising emergency examination of a person experiencing a mental health crisis [27]— in Far North Queensland (Australia). The key objective has been to investigate the temporal and spatial variation in relative risk of EE-O/As among postcodes of FNQ postcodes, based on more than 13000 unique EE-O/A admissions from 2009 to 2020, accounting for various exposure and confounding factors. To conduct this study we had to construct a bespoke dataset by combining disparate records of hospital, retail, demographic and various socio-economic predictors, aggregating all information at postcode level. To analyse the spatial and temporal variability of total EE-O/A in postcodes in a given year, a two-stage supervised probability modelling framework was used comprising Lasso regression for covariate selection and a spatial Conditional Autoregressive (CAR) model to integrate systematic environmental factors with spatial dependence (stochastic variation). Investigations reveal an increasing trend in EE-O/A admissions over the years corroborating previous publications[28–31], accompanied by evidence of positive spatial autocorrelation and association of EE-O/As with potential confounders and exposures. The distribution of EE-O/A admissions exhibited statistically significant clustering with geographically proximate postal regions that often had similar prevalent socio-economic conditions. Large spatial EE-O/A clusters were observed in the neighbourhood of three postal regions: Cairns (4870), Townsville (4810), and Mackay (4740), that also aligns with the region's three major regional hospitals and commercial districts. Statistical modelling established that the spatial distribution of socioeconomic and environmental characteristics of susceptible population, in these areas, aligned with the clustering of EE-O/A administered incidence, suggesting that underlying factors such as income, education, occupation, and retail density contribute to regional differences in EE-O/A incidences.

The CAR model identified that EE-O/A admissions were positively associated with the proportion of Aboriginal residents in a postal district (posterior mean = 0.351, 95% CI: 0.235–0.463) and with the number of pharmacies within a postal region (posterior mean = 0.422, 95% CI: 0.174–0.672). An increase in either of these factors was associated with a corresponding increase in EE-O/A counts. The association between Aboriginal population proportion and higher EE-O/A rates quantifies the ongoing disparities in mental health care access faced by Indigenous communities in regional Queensland [32]. Meanwhile, the positive association between EE-O/A rates and pharmacy density may reflect the tendency for pharmacies to be co-located with or near emergency services, thus, correlating with urban centres where EE-O/A reporting is concentrated. This association could facilitate future policy interventions studies that could harness pharmacies' expertise in reducing EE-O/A burden on emergency hospital departments. These could include calibrating opening hours for designated pharmacies.

**Table 3**
Bayesian estimation of socio-economic confounding factors significantly associated with EE-A/O admissions, based on model 37.

| Variable | Mean | SD | 2.5% Quant | Median | 97.5% Quant | Mode | KLD |
|---|---|---|---|---|---|---|---|
| (Intercept) | -0.530 | 1.746 | -3.923 | -0.545 | 2.944 | -0.544 | 0 |
| Aboriginal | 0.351 | 0.058 | 0.235 | 0.352 | 0.463 | 0.352 | 0 |
| Average Income | -0.697 | 0.252 | -1.196 | -0.695 | -0.205 | -0.695 | 0 |
| White Collar Occupation | -0.537 | 0.081 | -0.695 | -0.537 | -0.378 | -0.537 | 0 |
| Pharmacy Count | 0.422 | 0.126 | 0.174 | 0.421 | 0.672 | 0.421 | 0 |

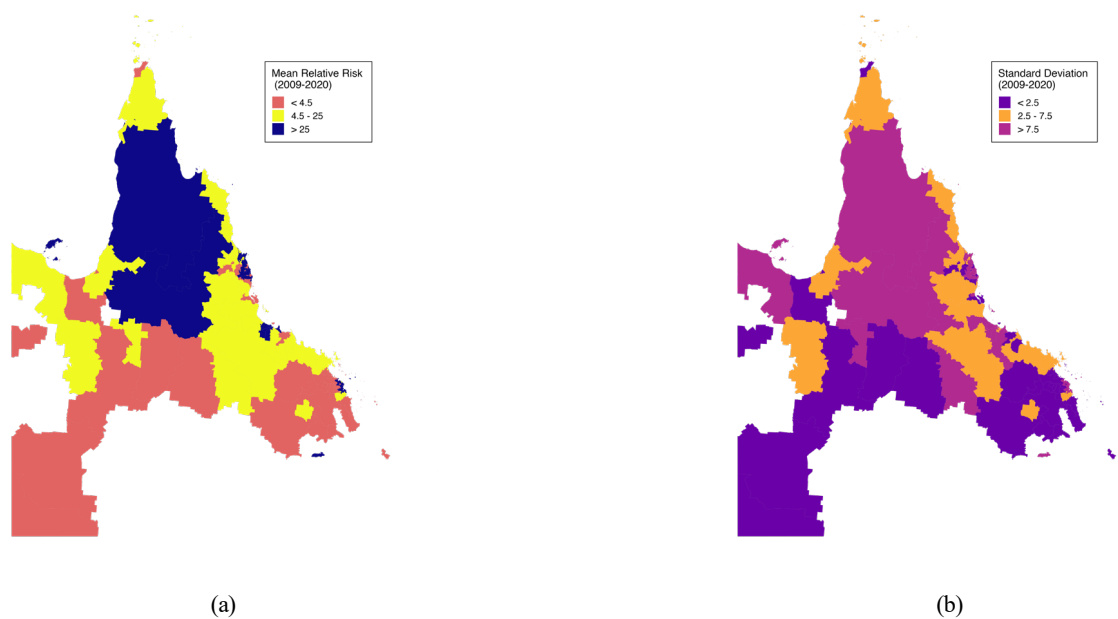

(a)  (b)

**Figure 5:** a) The choropleth shows the mean relative risk of EE-O/A from 2020 and, b) corresponding standard deviation

Economic status, such as average income, plays a critical role in determining the living conditions of people and access to health care services [33]. In this study, average income was found to be negatively associated with EE-O/A admissions (posterior mean = −0.697, 95%CI = −1.196, −0.205), suggesting that postcodes with higher income levels experienced fewer EE-O/A admissions while whereas lower-income areas exhibited higher rates. Furthermore, given the intrinsic link between income and occupational status, the study also identified a negative association between EE-O/A admissions and the prevalence of white-collar occupations. Individuals employed in white-collar roles are generally associated with higher incomes and more stable socioeconomic conditions, which may reduce their risk of mental health crises requiring emergency intervention. These findings highlight the relevance of identifying and addressing socioeconomic determinants of EE-O/A ad- missions. The pattern revealed that individuals residing in socioeconomically disadvantaged areas may face greater stressors, reduced access to mental health support, and consequently a higher likelihood of being subjected to EE-A/O interventions.

The spatial clustering, (Figure 3a) and (Figure 3b), of EE-O/A around major hospital centres -accounting for susceptible population- also indicate a shortage of required health-services in higher population density zones.

An examination of the five leading postal regions with highest relative risk, from 2009 to 2020 revealed patterns suggest that this could be a result of shifting policy. Cairns (postcode 4870) exhibited a marked increase in EE-O/A admissions in 2017, exceeding that of any other region, as illustrated in (Fig 6). Under the new regime, both public health patients and mental health patients are now all conflated and directed to Cairns Hospital emergency department as primary destination. This can also be seen in the fact that Moran's I statistic , depicted in Table 3a, shows significant positive-cluster trend since 2017 if we include the hospital postcodes while significance of stochastic spatial clustering disappears as we remove the hospital postcodes (see Appendix G). The consistently pronounced spatial auto correlation observed following the regulatory transition also suggests that the policy change may have contributed to spatial heterogeneity and geographic disparities, with postal regions having higher concentration of emergency departments being therefore more likely to experience an increased influx of patients under the EE-O/A framework from 5 March 2017. Consequently, there is a possibility that the policy change would have led to over-reporting into hospital emergency departments confounding reported residential postcodes with postcodes of hospitals. An unwitting consequence of such misrepresentation of postcodes is that besides systematic transition of clinical care burden it constrains epidemiology and hence hinders advancing appropriate population health care through prevention. But statistical validation of these reporting issues would require a separate study design.

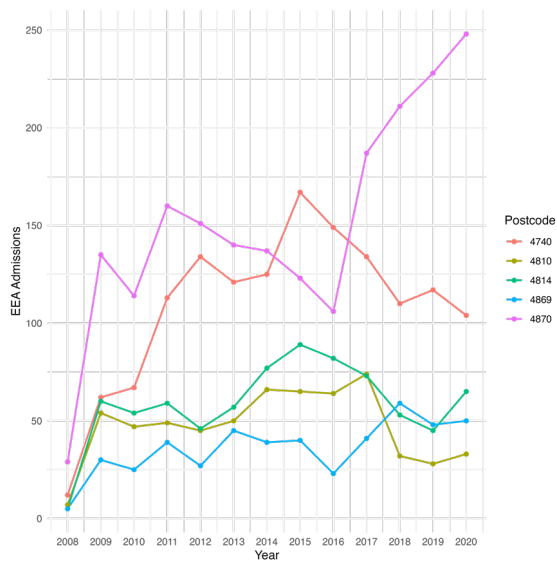

**Figure 6:** The graph illustrates the five postcodes that exhibited highest number of EE-O/A admissions out of 83 postcodes analysed from 2009 to 2020.

An additional confounder is that since 2017, Cairns also recorded elevated EE-O/A risk among Indigenous populations, shown Figure 8 in Appendix F) compared to non- Indigenous groups (shown Figure 9 in Appendix F), with Indigenous admissions surpassing non-Indigenous admissions in 2019 and 2020.

These spatial and temporal patterns point to differential policy impacts across demographic groups and suggest a disproportionate burden of EE-O/As among Indigenous communities. The spatial heterogeneities of social determinants combined with the policy shift is making it harder for Indigenous communities of FNQ, to receive requisite specialised care, but also appropriate lifestyle assistance, within an already constrained system. This is particularly troubling given that, according to the Australian Institute of Health and Welfare [34], approximately 3 in 10 Indigenous people reported experiencing severe levels of psychological distress in 2018-2019. Furthermore, postcode 4870 also showed an increased rate of EE-O/A admissions among individuals under the age of 18, highlighting a concerning trend in ever younger persons presenting to EDs overall. This in turn suggests significant burdens on the current health infrastructure and support systems and limits on accommodating the growing and diverse demands of EE-O/A episodes.

QPRIME's handling of EE-O/A incident data since 2009 has ensured high-quality records, supporting not only enforcement and investigations but also a comprehensive EE-O/A dataset. Thus, our analyses considered a much larger EE-O/A dataset compared to any other prior studies. Further- more, we collected and aggregated data from various sources to create an integrated repository. Our primary information sources included demographic, health and emergency, general practitioners, pharmaceutical, and retail-registration databases. While much of published literature on mental health informatics relies on hospital-based studies or web-scraped social media texts , this research developed a systematic, but painstaking, data integration framework allowing us to investigate the association of risk with quite disparate socio-economic factors. We did not find any published precedence of such broad data-fusion exercise in Australian mental health epidemiology in any of its States or Territories. For some time, various authors have argued for such data integration framework across the globe. However, as in all observational epidemiological studies that are dependent on public data, ecological bias and missing values pose challenges. A more detailed analysis of uncertainties of the data integration is left as a separate exercise.

However, our analyses are constrained by the availability and reliability of comprehensive data. A significant portion of the dataset was obtained from publicly accessible sources, while some were extracted from various health agencies. The use of 2021 census data to represent socioeconomic and demographic conditions across the entire study period (2009–2020) may not accurately reflect temporal changes or maintain consistent relationships over time. Furthermore, the data fusion process introduces the risk of ecological bias [35], potentially challenging the interpretation of results. The study also did not conduct a comparative analysis of pre- and post-implementation periods of the emergency examination provisions, as this would necessitate a redefinition of the outcome variable, such as examining the time interval between admissions or evaluating outcomes relative to the policy implementation date. However, such a distinction is critical, as the duration of patient support significantly impacts health outcomes and could provide valuable insights into the policy's effectiveness.

## 6. Conclusion

The EE-O/A dataset provides valuable information on the legal challenges and systemic issues faced by mental health patients in Queensland. We investigate the spatial and temporal dynamics of EE-O/A incidences by establishing a novel framework involving comprehensive data collection, processing, and assimilation into a single mental health database. The study aims to provide researchers and policy-makers with insights into the patterns, trends, and systemic determinants of EE-A/O occurrences to support effective, evidence-based intervention strategies, crisis management, and resource allocation.

Future research would build on this framework by incorporating health officer– (ambulance) driven EE-A/O

events to capture a wider spectrum of first responders' response and identify differential patterns and trends. Eventually, the work can be extended to other states, thereby creating state-wide models and facilitating comparative analyses with other jurisdictions in Australia where mental health legislation falls under different governing bodies. This would advance the understanding of systemic health challenges nationwid

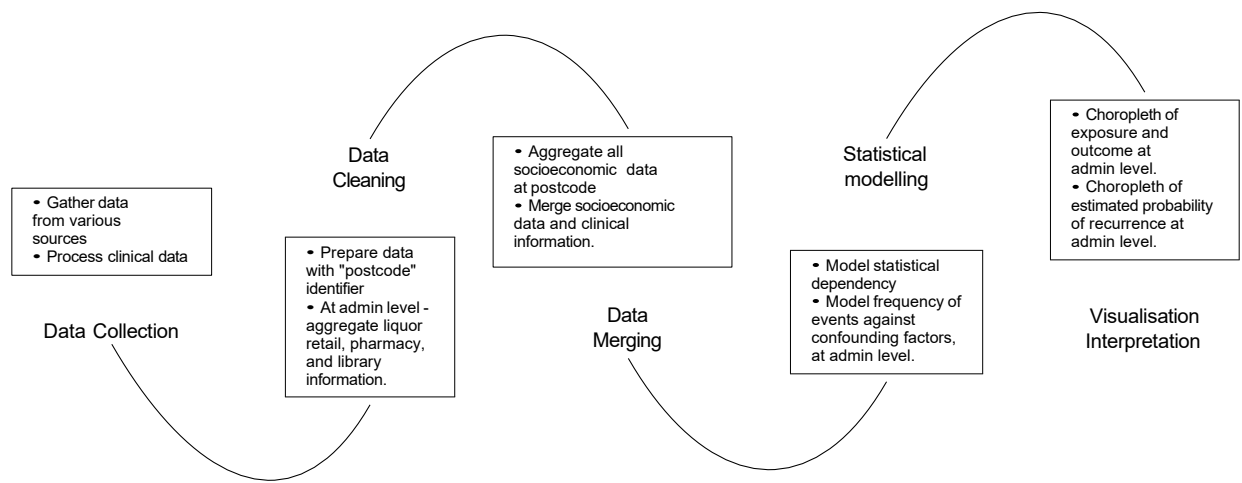

**Figure 7:** Flowchart illustrating the workflow of data collection, aggregation, and subsequent statistical analysis used in the study.

**Table 4**
The table represents data of potential association of socio-economic characteristics and EE-A/O admissions of the five postal regions with highest recorded admissions from 2009 to 2020.

| Postcode | EE-O/A Admissions | Non-Indigenous | Aboriginal | Nil Income | Average Income | Blue-Collar Jobs | Liquor Stores |
|---|---|---|---|---|---|---|---|
| 4870 | **1951** | 59600 | **3422** | 3736 | 1106 | 15247 | **341** |
| 4740 | 1406 | **73200** | 3416 | **4519** | 1217 | **21835** | 150 |
| 4814 | 766 | 38959 | 2575 | 2647 | 1097 | 10416 | 29 |
| 4810 | 610 | 18650 | 892 | 941 | 1299 | 4411 | 154 |
| 4869 | 471 | 15945 | 1688 | 1109 | 1030 | 4883 | 12 |

and inform policies aimed at improving access, care, and intervention for mental health issues.

## A. Appendix: Data aggregation and modelling framework

The Fig 7 and represents the data processing, aggregation and statistically modelling framework.

## B. Appendix: Social-Economic Characteristics

## C. Appendix: Variables for modelling

The following outlines the operational definitions of the spatially referenced variables derived from the data sources used in this study:

i. **Number of Distinct EEA Admissions**: Emergency Examination Authority (EEA) admissions were aggregated by postcode and year. This variable is modelled as the response variable, denoted as **Y**, representing the count of EEA admissions within a given postcode in a specific year.

ii. **Australian Bureau of Statistics Variables**:

(a) *Demographic Profile*: categorised into five groups, namely, Non-Indigenous, Aboriginal, Torres Strait Islander, Both, or Not Stated.
(b) *Age Distribution*: age was summarised into three categories and aggregated at the postcode level: <18 years, 18–64 years, and ≥65 years, in accordance with previous research [36].
(c) *Education Level*: likewise, the highest year of school data was categorised into: < year 8; year 9-12 and school not attended
(d) *Occupation*: occupations were classified into two broad categories, blue-collar and white-collar, based on the framework by [37]. Blue-collar occupations include technicians and trades workers, community and personal service workers, and labourers. White-collar occupations include man- agers, clerical and administrative staff, and sales workers. The distinction is based on the nature of work, with blue-collar roles being primarily physical and white-collar roles primarily administrative or mental.
(e) *Economic Status*: categorised into Negative Income (loss), Nil Income (zero), and Weekly Average Income. Weekly average income was estimated by assigning the median value of each income bracket, multiplying it by the corresponding frequency, and dividing by the adjusted population, which excludes individuals with nil or negative income. The weekly income from ABS was recorded at intervals. Instead of using all different intervals, the average weekly income was calculated by taking the sum of the median of the income interval multiplied it by the corresponding frequency (number of individuals within that range) and dividing by adjusted population, as shown in Equation 9. For example, the range "$1 - $149" is assigned a mid-point of $75 and this is to ensure uniform distribution within each range. The adjusted population excludes individuals for whom income was not stated, as well as those reporting nil or negative income. Table 5 shows the new columns in the weekly income variable extracted from ABS.

| postcode | negative | nil | average |
|---|---|---|---|
| 4000 | 76 | 1733 | 1344 |
| 4005 | 65 | 676 | 1691 |
| 4006 | 58 | 1017 | 1431 |
| 4007 | 49 | 970 | 1605 |
| 4008 | 0 | 12 | 1175 |

**Table 5**
A detailed method employed to estimate an average income across postcodes from the weekly income intervals from ABS

iii. **Number of Liquor Stores**: represents the count of liquor stores in each postcode whose primary function is involved direct sale to consumers. Any other categories where the sale of liquor was a secondary purposed were excluded.

iv. **Number of Schools and Libraries** denotes the total number of schools and libraries within each postcode.

v. **Number of Pharmacies and General Practitioners**: it is total number of pharmacies and general practitioner clinics located within each postcode.

The missing values for liquor stores, schools, libraries, pharmacies, and GPs within each postcode had imputed using a population-based negative binomial generalised linear model as population often determines the availabilities of these services and infrastructure.

## D. Appendix: Bayesian Parameter Estimation

Table 6 shows the parameter estimations of the Lasso selected variables after fitting to **Model 3**.

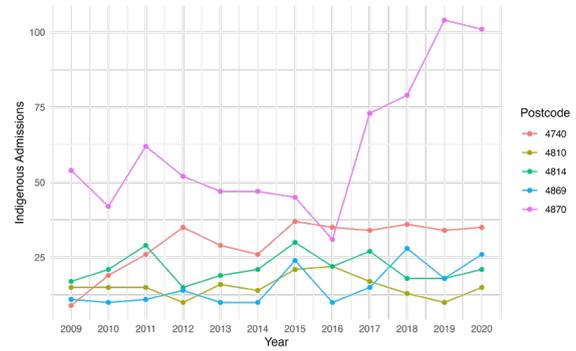

**Figure 8:** Trend of indigenous population admissions for the five postal regions with the highest recorded EE-A/O admission from 2009 to 2020.

## E. Appendix: Model Equation

Assuming EEA follows Negative Binomial Distribution, the response variable $Y$, was modelled using Conditional Autoregressive model as follows:

$$y_{it} \sim \text{NegBin}(\mu_{it}, \theta),$$

$$\log(\mu_{it}) = \beta_0 + \sum_{j=1}^{p} \beta_j \log(x_{ijt}) + s_i + \gamma_t$$

$$i = 1, 2, \ldots, N; \quad t = 1, 2, \ldots, T$$

where $y_{it}$ is the observed EEA count for postal region $i$ at time $t$; $\mu_{it}$ is the expected count; and $\theta$ is the dispersion parameter of the Negative Binomial distribution. The term $\beta_0$ represents the intercept, while $\beta_j$ are the regression coefficients corresponding to the log-transformed covariates $x_{ijt}$. The spatially structured random effect for region $i$ is denoted by $s_i$, which is commonly modelled using a Conditional Autoregressive (CAR) prior. Temporal trends are captured by $\gamma_t$, the temporally structured random effect for time $t$.

## F. Appendix: Indigenous and Non-Indigenous Admissions

To understand the differential impact of policy change, we analysed the EE-O/A admissions of indigenous and non-indigenous admissions from 2009 to 2020. Fig 8 and Fig 9 shows the trends of EE-O/A admissions in the respective demographic groups.

## G. Appendix: Moran's I without Hospital Postcodes

To understand the impact of the policy change on spatial clustering, we calculated the Global Moran's I statistics with postal regions with hospitals removed from the dataset. Table 8 shows the spatial autocorrelation for each year from 2009 to 2020 and the aggregated Moran's I for the full dataset without hospital postcodes.

**Table 6**

A conditional autoregressive model of potential association of socioeconomic, environmental, and retail factors with of EE-A/O admissions

| Variable | Mean | SD | 2.5% Quant | Median | 97.5% Quant | Mode | KLD |
|---|---|---|---|---|---|---|---|
| (Intercept) | -0.133 | 1.742 | -3.527 | -0.145 | 3.327 | -0.145 | 0 |
| **Aboriginal** | **0.382** | **0.073** | **0.234** | **0.384** | **0.523** | **0.384** | **0** |
| School Not Attended | 0.022 | 0.081 | -0.136 | 0.021 | 0.183 | 0.021 | 0 |
| **Average Income** | **-0.748** | **0.258** | **-1.257** | **-0.748** | **-0.243** | **-0.748** | **0** |
| **White Collar Occupation** | **-0.558** | **0.085** | **-0.725** | **-0.558** | **-0.392** | **-0.558** | **0** |
| **School Count** | **-0.226** | **0.112** | **-0.443** | **-0.227** | **-0.004** | **-0.227** | **0** |
| Liquor Count | 0.074 | 0.065 | -0.054 | 0.074 | 0.203 | 0.074 | 0 |
| **Pharmacy Count** | **0.423** | **0.153** | **0.122** | **0.423** | **0.723** | **0.423** | **0** |

**Table 7**

The table illustrates Lasso Regression estimation of the coefficients of the potential confounding factors.

| Variable | Coefficient |
|---|---|
| Intercept | 1.11 |
| Aboriginal Population | 0.50 |
| Not attended School | 0.05 |
| Average Income | -0.49 |
| White Collar Occupation | 0.18 |
| School count | -0.008 |
| Library count | 0.11 |
| Pharmacy Count | 0.28 |

**Table 8**

Annual Global Moran's I Spatial Autocorrelation of EE-O/A Admissions (2009–2020), and Aggregated Moran's I (2009–2020) by Postal Region, Excluding Regions Containing Hospitals and Emergency Departments."

| Year | P-value | Z-score | Moran's I |
|---|---|---|---|
| 2009 | **0.004** | 2.67 | 0.222 |
| 2010 | 0.247 | 0.653 | 0.050 |
| 2011 | 0.671 | -0.444 | -0.0566 |
| 2012 | 0.553 | -0.133 | -0.027 |
| 2013 | 0.057 | 1.58 | 0.115 |
| 2014 | 0.170 | 0.954 | 0.062 |
| 2015 | 0.144 | 1.064 | 0.069 |
| 2016 | 0.198 | 0.850 | 0.053 |
| 2017 | **0.016** | 2.146 | 0.164 |
| 2018 | 0.265 | 0.628 | 0.037 |
| 2019 | 0.183 | 0.905 | 0.054 |
| 2020 | 0.115 | 1.196 | 0.094 |
| aggregated | **0.011** | 2.272 | 0.151 |

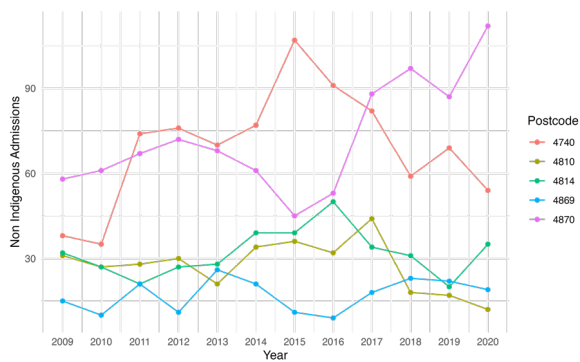

**Figure 9:** Trend of non-indigenous population admissions for the five postal regions with the highest recorded EE-A/O admission from 2009 to 2020.

## References

[1] Malcolm Smith. *Medical Law*. LexisNexis Butterworths, 2020.

[2] Ian Freckelton. Mental health law. In Fiona McDonald, Ben White, and Lindy Willmott, editors, *Health Law in Australia*, pages 729–772. Lawbook Co., Thomson Reuters (Professional) Australia, 3 edition, 2018.

[3] Public health act 2005 (qld). Queensland Government, 2005. Queensland legislation.

[4] Queensland Government. Parliamentary debates, public health act 2005 (qld). Hansard, 2005. Queensland Parliament.

[5] Coroners Court of Queensland. Recommendations from inquest: Inquest into the deaths of anthony william young, shaun basil kumeroa, edward wayne logan, laval donovan zimmer, and troy martin foster. Findings of Mr Terry Ryan, State Coroner, Coroners Court of Queensland, 2017. URL https://www.courts.qld.gov.au/__data/assets/pdf_file/0003/544045/cif-young-kumeroa-logan-zimmer-foster-20171214.pdf. COR 2013/2988; 2014/3598; 2014/4321; 2014/4239; 2014/4357. Delivered 20 October 2017 at Brisbane.

[6] Australian Bureau of Statistics. Snapshot of queensland, 2024. URL https://www.abs.gov.au. Accessed: 2024-09-18.

[7] A Rogier T Donders, Geert JMG Van Der Heijden, Theo Stijnen, and Karel GM Moons. A gentle introduction to imputation of missing values. *Journal of clinical epidemiology*, 59(10):1087–1091, 2006.

[8] Queensland Government. Open data portal, 2024. URL https://www.data.qld.gov.au/. Accessed: 2024-9-19.

[9] Queensland Government, Department of Health. Pharmacy business ownership administration system, 2024. URL https://www.pboas.health.qld.gov.au/. Accessed: 2024-10-16.

[10] Australian Commission on Safety and Quality in Health Care. Australian commission on safety and quality in health care. Australian Government Department of Health and Aged Care, May 2025. URL https://www.safetyandquality.gov.au/. Accessed: May 22, 2025.

[11] P. A. P. Moran. Notes on continuous stochastic phenomena. *Biometrika*, 37:17–23, 1950. doi: 10.1093/biomet/37.1-2.17. URL https://doi.org/10.1093/biomet/37.1-2.17.

[12] Jared Aldstadt. Spatial clustering. In *Handbook of applied spatial analysis: Software tools, methods and applications*, pages 279–300. Springer, 2009.

[13] Hongfei Li, Catherine A Calder, and Noel Cressie. Beyond moran's i: testing for spatial dependence based on the spatial autoregressive model. *Geographical analysis*, 39(4):357–375, 2007.

[14] Robert Tibshirani. Regression shrinkage and selection via the lasso. *Journal of the Royal Statistical Society Series B: Statistical Methodology*, 58(1):267–288, 1996.

[15] Jerome Friedman, Trevor Hastie, Rob Tibshirani, Balasubramanian Narasimhan, Kenneth Tay, Noah Simon, and Junyang Qian. Package 'glmnet'. *CRAN R Repositary*, 595:874, 2021.

[16] Peter J Diggle and Emanuele Giorgi. *Model-based geostatistics for global public health: methods and applications*. Chapman and Hall/CRC, 2019.

[17] Noel Cressie and Christopher K Wikle. *Statistics for spatio-temporal data*. John Wiley & Sons, 2011.

[18] J. Besag. Spatial interaction and the statistical analysis of lattice systems. *Journal of the Royal Statistical Society Series B: Statistical Methodology*, 36(2):192–225, January 1974. doi: 10.1111/j.2517-6161.1974.tb00999.x.

[19] P. Whittle. On stationary processes in the plane. *Biometrika*, 41(3/4):434–449, 1954. doi: 10.2307/2332724. URL https://doi.org/10.2307/2332724.

[20] Leonhard Knorr-Held and Julian Besag. Modelling risk from a disease in time and space. *Statistics in medicine*, 17(18):2045–2060, 1998.

[21] Julian Besag, Jeremy York, and Annie Mollié. Bayesian image restoration, with two applications in spatial statistics. *Annals of the institute of statistical mathematics*, 43(1):1–20, 1991.

[22] Håvard Rue, Sara Martino, and Nicolas Chopin. Approximate bayesian inference for latent gaussian models by using integrated nested laplace approximations. *Journal of the Royal Statistical Society Series B: Statistical Methodology*, 71(2):319–392, 2009.

[23] Finn Lindgren and Håvard Rue. Bayesian spatial modelling with r-inla. *Journal of statistical software*, 63:1–25, 2015.

[24] Haakon Bakka, Håvard Rue, Geir-Arne Fuglstad, Andrea Riebler, David Bolin, Janine Illian, Elias Krainski, Daniel Simpson, and Finn Lindgren. Spatial modeling with r-inla: A review. *Wiley Interdisciplinary Reviews: Computational Statistics*, 10(6):e1443, 2018.

[25] Marta Blangiardo and Michela Cameletti. *Spatial and spatio-temporal Bayesian models with R-INLA*. John Wiley & Sons, 2015.

[26] Sander Greenland. Model-based estimation of relative risks and other epidemiologic measures in studies of common outcomes and in case-control studies. *American journal of epidemiology*, 160(4):301–305, 2004.

[27] Queensland Government. Mental Health Act 2016, 2016. URL https://www.legislation.qld.gov.au/view/pdf/inforce/current/act-2016-005. Updated 1 July 2024; Accessed: 2024-08-28.

[28] Alan R Clough, Angela Evans, Veronica Graham, Janet Catterall, Richard Lakeman, John Gilroy, Gregory Pratt, Joe Petrucci, Ulrich Orda, Rajesh Sehdev, et al. Emergency examination authorities in queensland, australia. *Emergency Medicine Australasia*, 35(5):731–738, 2023.

[29] A. R. Clough et al. Recent amendments to queensland legislation make mental health presentations to hospital emergency departments more difficult to scrutinise. *Emerg Medicine Australasia*, 34(1):130–133, February 2022. doi: 10.1111/1742-6723.13878.

[30] T. Meehan and T. Stedman. Trends in the use of emergency examination orders in queensland since the implementation of the mental health intervention project. *Australas Psychiatry*, 20(4):287–290, August 2012. doi: 10.1177/1039856212450082.

[31] Sourav Das, Janet Catterall, Richard Stone, and Alan R Clough. "the reasons you believe...": An exploratory study of text driven evidence gathering and prediction from first responder records justifying state authorised intervention for mental health episodes. *Computer Methods and Programs in Biomedicine*, 254:108257, 2024.

[32] B. E. Kavanagh, K. B. Corney, H. Beks, L. J. Williams, S. E. Quirk, and V. L. Versace. A scoping review of the barriers and facilitators to accessing and utilising mental health services across regional, rural, and remote australia. *BMC Health Services Research*, 23(1):1060, October 2023. doi: 10.1186/s12913-023-10034-4.

[33] Darcy Jones McMaughan, Oluyomi Oloruntoba, and Matthew Lee Smith. Socioeconomic status and access to healthcare: interrelated drivers for healthy aging. *Frontiers in Public Health*, 8:231, 2020. doi: 10.3389/fpubh.2020.00231.

[34] Australian Institute of Health and Welfare. Mental health expenditure, 2023. URL https://www.aihw.gov.au/mental-health/topic-areas/expenditure. Accessed: 2024-10-09.

[35] S. Greenland and H. Morgenstern. Ecological bias, confounding, and effect modification. *International Journal of Epidemiology*, 18(1):269–274, March 1989. doi: 10.1093/ije/18.1.269. Erratum in: Int J Epidemiol. 1991 Sep;20(3):824.

[36] S. Das, J. Catterall, R. Stone, and A. R. Clough. "the reasons you believe...": An exploratory study of text driven evidence gathering and prediction from first responder records justifying state authorised intervention for mental health episodes. *Computer Methods and Programs in Biomedicine*, 254:108257, September 2024. doi: 10.1016/j.cmpb.2024.108257.

[37] X. Hu, S. Kaplan, and R. S. Dalal. An examination of blue- versus white-collar workers' conceptualizations of job satisfaction facets. *Journal of Vocational Behavior*, 76(2):317–325, April 2010. doi: 10.1016/j.jvb.2009.10.014.